\documentclass[superscriptaddress,twocolumn,amssymb,footinbibprl,aps]{revtex4-1}
\usepackage{graphicx,bm,times}
\usepackage{tabularx} 
\usepackage{lipsum}
\usepackage{amsmath}
\usepackage{amssymb}
\usepackage{gensymb}
\usepackage{bbold}
\usepackage{float}
\usepackage{color,soul, mathrsfs}
 \usepackage{relsize}

\usepackage[]{natbib}

\begin{document}

\title{
Sub-shot-noise transmission measurement enabled by optically gating on single photons
}

\author{J. Sabines-Chesterking}
\author{R. Whittaker}
\affiliation{Quantum Engineering Technology Labs, H. H. Wills Physics Laboratory and Department of Electrical \& Electronic Engineering, University of Bristol, BS8 1FD, UK.}
\author{S. K. Joshi}
\affiliation{Institute for Quantum Optics and Quantum Information (IQOQI)\\
Austrian Academy of Sciences, Boltzmanngasse 3, A-1090 Vienna, Austria}
\author{P. M. Birchall}
\author{P. A. Moreau}
\author{A. McMillan}
\author{H. V. Cable}
\author{J. L. O'Brien}
\author{J. G. Rarity}
\author{J. C. F. Matthews}
\affiliation{Quantum Engineering Technology Labs, H. H. Wills Physics Laboratory and Department of Electrical \& Electronic Engineering, University of Bristol, BS8 1FD, UK.}

\begin{abstract}
Harnessing the unique properties of quantum mechanics offers the possibility to deliver new technologies that can fundamentally outperform their classical counterparts. These technologies only deliver advantages when components operate with performance beyond specific thresholds. For optical quantum metrology, the biggest challenge that impacts on performance thresholds is optical loss. Here we demonstrate how including an optical delay and an optical switch in a feed-forward configuration with a stable and efficient correlated photon pair source reduces the detector efficiency required to enable quantum enhanced sensing down to the detection level of single photons. When the switch is active, we observe a factor of improvement in precision of 1.27 for transmission measurement on a per input photon basis, compared to the performance of a laser emitting an ideal coherent state and measured with the same detection efficiency as our setup. When the switch is inoperative, we observe no quantum advantage. 
 
\end{abstract}
\date{\today}
\maketitle

Quantum mechanics quantifies the highest precision that is achievable in each type of optical measurement~\cite{gardiner2004quantum,schnabel2010quantum, gi-nphot-5-222}. Single photon probes measured with single photon detectors are in principle optimal for gaining the most precision per-unit intensity when measuring optical transmission~\cite{Adesso09}. However, in practice, optical loss and low component efficiencies prevent an advantage from being achieved using single photon detectors~\cite{thomas2011real}. One way to reduce the impact of lower component efficiency is to incorporate fast optical switching and an optical delay with schemes that are based on heralded generation of quantum sates~\cite{Jakeman86}. 
This then enables use of a quantum state conditioned on the successful detection of a correlated signal --- this is referred to as feed-forward.

\begin{figure}
 \centering
\includegraphics[width=.9\columnwidth]{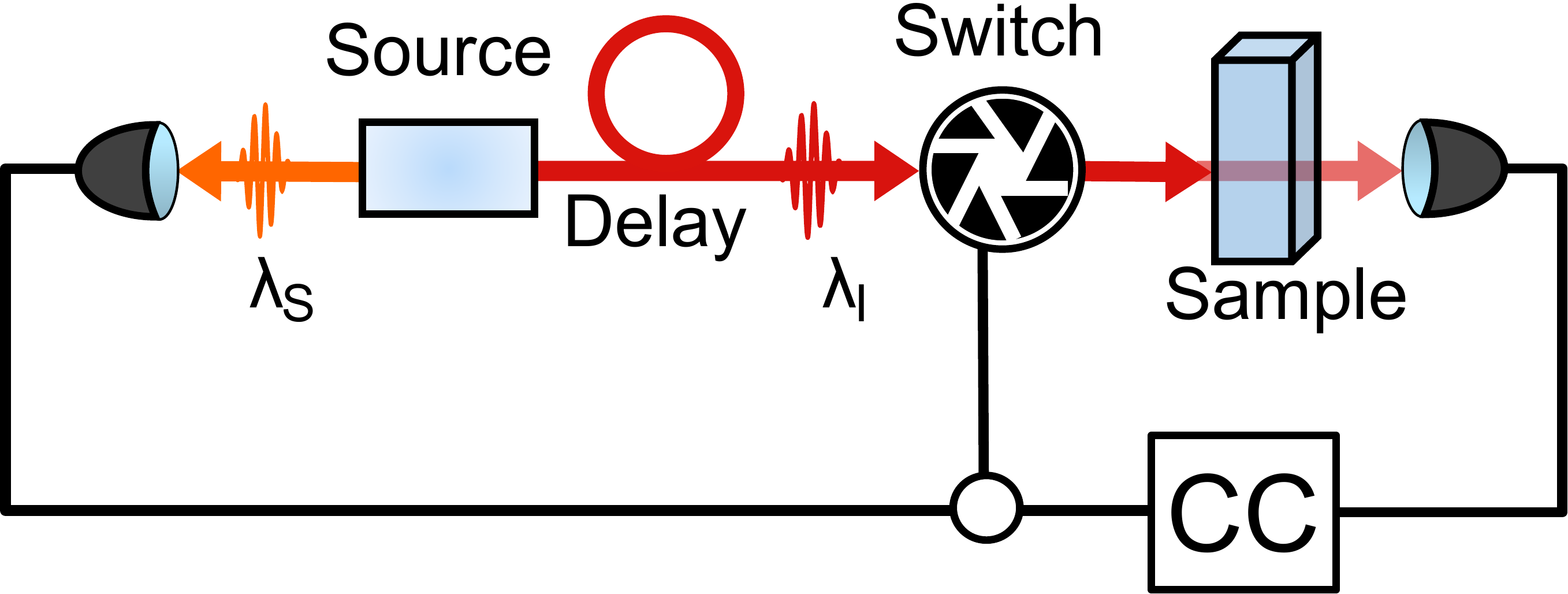}
\caption{\textbf{Photon pair feed-forward transmission measurement.} Photon pairs of signal ($\lambda_S$) and idler ($\lambda_I$) photons are simultaneously emitted into two channels. Once a signal photon is detected, it opens a switch in the idler photon's channel to allow probing of a sample with the idler photon. The transmission estimate is obtained from the ratio of the number of coincidence detection (CC) and signal photon detection events.}
\label{scheme}
\end{figure}

Feed-forward is key for demonstrations of optical quantum computing~\cite{Prevedelfeedforward}, it has been used in experiments that increase the generation rate~\cite{ma2011experimental,sydneymultiplex, ka-optica-2-1010, mendoza2016active,fr-arxiv:1603.06260} and signal-to-noise ratio~\cite{brida2012extremely} of heralded single photons, it has been used to calibrate single photon detectors~\cite{1612-202X-3-3-001} and it has also been applied to gather evidence of single photon sensitivity in animal vision~\cite{frogexperiment}. 
Jakeman and Rarity proposed in Ref.~\cite{Jakeman86} using feed-forward with correlated photon pairs to enable sub shot noise optical transmission measurements when component efficiency is otherwise not sufficient to permit a quantum advantage in passive direct detection~\cite{heidmann1987observation, rarity1987observation, br-nphot-4-227}. But despite becoming identified as key to more general multi-photon entangled quantum state engineering for quantum metrology~\cite{ca-prl-99-163604,mccusker2009efficient}, feed-forward has not been implemented for quantum enhanced parameter estimation. Here we implement the proposal featured in Ref.~\cite{Jakeman86} (Fig.~\ref{scheme}) to realise sub shot noise measurement of transmissitivity, using single photon detectors that are too low in efficiency to enable sub shot noise performance in a passive measurement.

 \begin{figure}
  \centering
\includegraphics[width=.9\columnwidth]{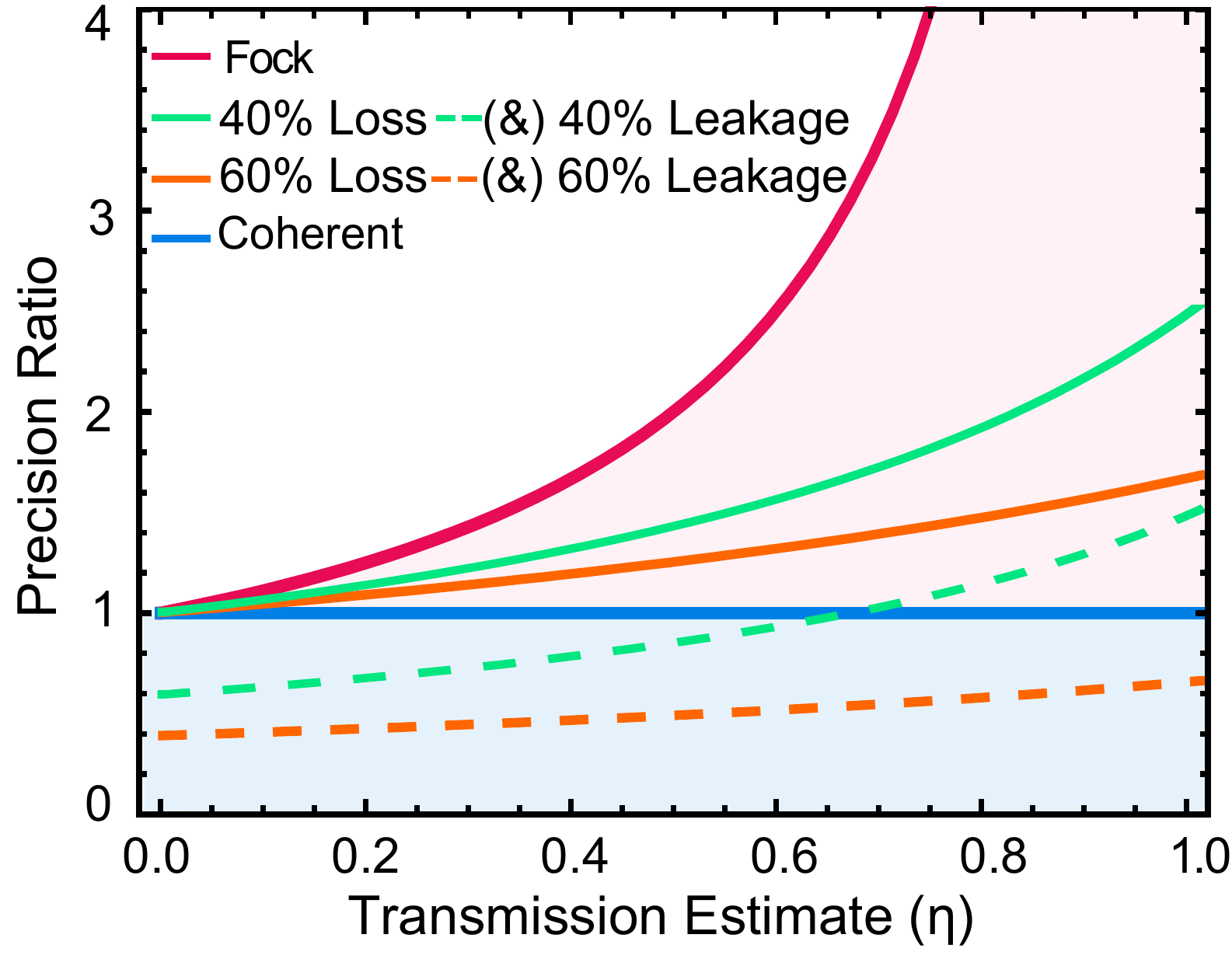}
\caption{\textbf{Theoretical performance of the photon pair feed-forward transmission measurement.} Precision achievable relative to coherent states is plotted
as a function of sample transmission $\eta$ for
 an average input intensity of $\bar{N}=1$ photons. The pink curve represents the ideal case of a heralded Fock state with no setup loss ($\eta_\textrm{det}.\eta_\textrm{source} = 1$), aside from the sample's transmission, and no switch leakage ($\eta_S = 1$). Illustrating the effect of experimental imperfections, the green and orange solid curves correspond to mixed states with setup losses $\eta_\textrm{det}.\eta_\textrm{source}=0.4$ and $\eta_\textrm{det}.\eta_\textrm{source}=0.6$ and with $\eta_S = 1$. The green and orange dashed curves represent respectively performance with $\eta_\textrm{det}.\eta_\textrm{source}=0.4$ and leakage $1-\eta_{S}=0.4$, and $\eta_\textrm{det}.\eta_\textrm{source}=0.6$ and $1-\eta_{S}=0.6$. The blue curve represents the shot noise limit. The light pink region reflects the area where there is a quantum advantage.}
\label{FIgraph}
\end{figure}

The transmissivity $\eta$ of a sample is in general estimated by measuring the reduction of light intensity from a known mean input value $\bar{ N}_{\text{in}}$, to a reduced mean value $\bar{N}_{\text{\text{out}}}$ according to $\eta=\bar{N}_{\text{in}}/\bar{N}_{\text{out}}$. The precision with which $\eta$ can be measured is dependent on the type of light  used to probe the channel. When estimating $\eta$ with an ideal coherent state probe $|\alpha\rangle$, the precision will be given by $(1/\Delta^2\eta)_{\alpha} = {\nu\bar{N}_{\text{in}}/\eta}$, where $\bar{N}_{\text{in}}$ is the average number of probe photons and $\nu$ is the number of repetitions of the measurement. This is the shot-noise limit and it is the upper-bound on the precision achievable with classical measurements~\cite{gi-nphot-5-222}. Higher precision can therefore be achieved by increasing the input intensity and the number of repetitions. 
For a fixed intensity and fixed number of repetitions $\nu$, non-classical states of light can provide an enhancement in precision over coherent state probes. The photon number probability distribution of a Fock state of $\bar{N}_{\text{in}}=N_{\text{in}}$ photons after passing through a lossy channel follows the Binomial distribution $P(N_{\text{out}},N_{\text{in}},\eta)=\binom{N_{\text{in}}}{N_{\text{out}}}\eta^{N_{\text{out}}}(1-\eta)^{N_{\text{in}}-N_{\text{out}}}$.  
So for fixed $\bar{N}_{\text{in}}$ and $\nu$, the Fock state probe achieves a higher precision than the coherent state~\cite{Adesso09}:
\begin{equation}
(1/\Delta^2 \eta)_{F} ={\nu \bar{N}_{\text{in}}}/{\eta (1-\eta)} > (1/\Delta^2\eta)_{\alpha}.
\label{deltaetaF}
\end{equation}

\begin{figure*}
  \centering
 \includegraphics[width=0.90\textwidth]{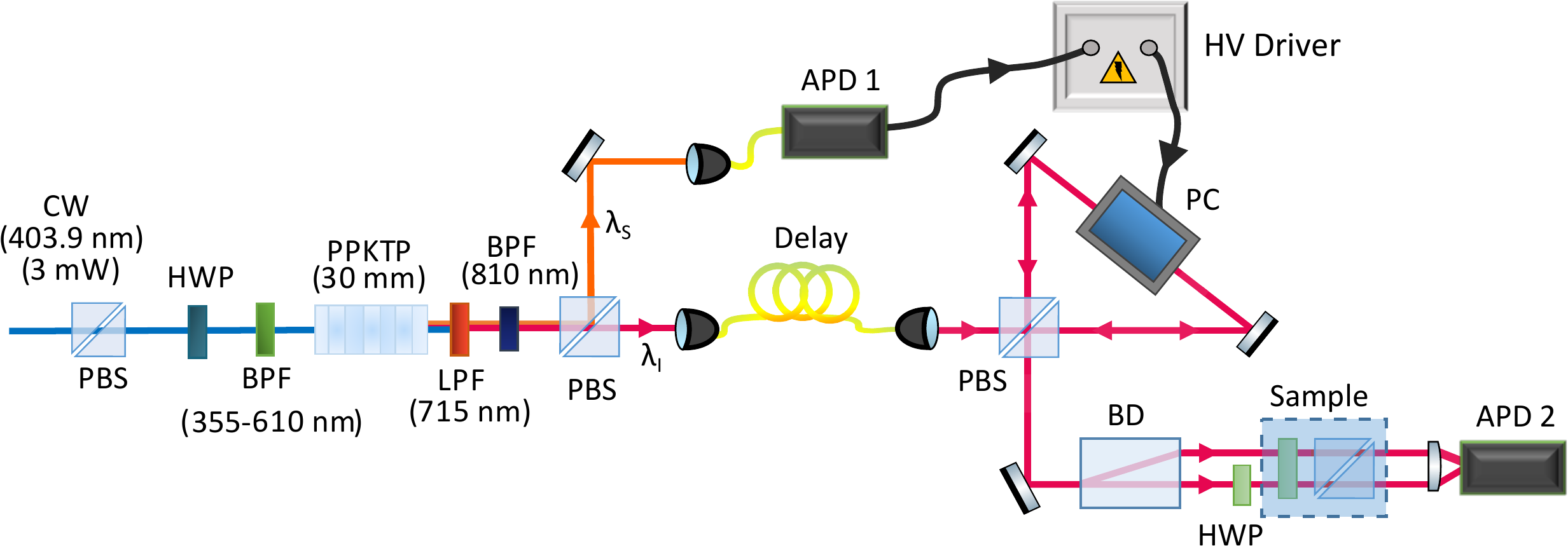}
 \caption{\textbf{Experimental setup.} Photon pairs are generated via type II collinear SPDC using a 30 mm PPKTP crystal pumped with a continuous wave (CW) 404 nm laser. While the idler photon goes through a delay line, the signal photon is detected and triggers an optical switch. The switch is a commercially available free-space Pockels cell type modulator consisting of two lithium niobate crystals inside a Sagnac loop. The source was mounted in a cage system to reduce vibrational noise.}
 \label{setup}
\end{figure*}

The performance of Fock states can be accessed by using correlated photon pairs generated from a spontaneous parametric down conversion process (SPDC). Signal photons of each correlated pair are sent directly for detection to herald the presence of the corresponding idler photon which is used to probe a sample. The transmissivity of the photons through the sample can then be estimated from the Klyshko (heralding) efficiency~\cite{klyshko1980use} of the idler channel $\eta_I$,  which  is the ratio of the number of photons detected coincidentally across the two channels $N_C$ and the total number of detected herald (signal) photons $N_S$: $\eta_{I} = \bar{N}_{C}/\bar{N}_{S}$.

To obtain a quantum advantage using the Klyshko efficiency as the transmission estimator it is required to have a strong correlation between the number of signal and idler photons such that the difference between the coincidence and the signal count is due only to the absorption of the sample~\cite{heidmann1987observation, rarity1987observation, br-nphot-4-227}.
This is generally not the case when there is loss in either the signal or the idler channel, so when system performance prohibits having a high correlation, one can selectively analyse subsets of recorded data in post-selection to observe sub-shot noise behaviour \cite{whittaker2015quantum}. However, in practice, a sample measured with post-selected events will be over-exposed, with photons that are unaccounted for due to lost counterpart heralding  photons. This results in a strategy that performs worse than using a coherent state when analysis is normalized to per input probe photon. By introducing an optical switch into the setup, as sketched in Fig.~\ref{scheme}, which only allows photons incident on a sample when a signal photon has been detected we increase the level of correlation, suppressing the detrimental effect of loss in the signal channel. However, there are still two main mechanisms that degrade the performance of the photon pair strategy using a switch. The first is loss of the idler photon in the photon source $1-\eta_{\textrm{source}}$ and at the detector  $1-\eta_{\textrm{det}}$, which together with sample transmission $\eta$ redefines the Klyshko efficiency $\eta_I= \eta_{\textrm{source}} . \eta .\eta_{\textrm{det}}$. For a single photon Fock state, ${\rho}=|1\rangle\langle 1|$, $\eta_I$ modifies the state according to
\begin{equation}
\rho\rightarrow {\rho}'=(1-\eta_{I})|0\rangle\langle0|+\eta_{I}|1\rangle\langle 1|,
\label{mix}
\end{equation}
which still follows a (sub-Poissonian) Binomial distribution and therefore still outperforms coherent states per input photon. But as loss increases $\eta_{\textrm{source}}.\eta_{\textrm{det}}\rightarrow0$, the measured photon number distribution tends towards Poissonian. The second degradation mechanism is imperfect optical switching that leaks unheralded photons through the sample. 
We plot examples of the effect of both of these mechanisms in 
Fig.~\ref{FIgraph}, in terms of the ratio between the precision achievable using a Fock state that has either been degraded by loss or incorrectly heralded with switch leakage, denoted $1/(\Delta^2\eta)_{F'}=~\eta_S/{\eta(1-\eta\eta_I)}$, and the precision achievable with a coherent state probe  $1/(\Delta^{2}\eta)_{\alpha}=1/\eta$ with the same detector efficiency. This ratio is a figure of merit that determines when a quantum advantage is obtained---that is when $R=\eta_S/(1-\eta.\eta_I)>1$. Note that this expression leads to the condition found in Ref.~\cite{Jakeman86} where it was shown that for obtaining a quantum advantage over using a coherent state it is necessary that
\begin{equation}
\eta_I+\eta_S>1.
\label{JakemanCondition}
\end{equation}

The experimental setup we used to implement feed-forward transmission measurement is shown in Fig.~\ref{setup}. Photon pairs were generated via collinear type II SPDC using a periodically-poled potassium titanyl phosphate crystal (PPKTP), pumped with a continuous wave (CW) laser diode ($\lambda_p=403.9$ nm) and spectrally tuned by controlling its temperature. The wavelengths of the signal and idler photons were $\lambda_s=792$~nm and $\lambda_i=824$~nm, each with a spectral width of $\pm$0.4~nm. After down conversion the pump was removed using a 715~nm long-pass filter (LPF) and a 50~nm wide bandpass filter (BPF) centered at 808 nm. Photon pairs were split deterministically using a polarisation beamsplitter (PBS), sending the idler photon through the delay line while the correlated signal photon was collected with a single mode fibre and detected using an avalanche photodiode (APD). 

The detected signal photon triggered an optical switch implemented with a Pockels cell modulator composed of two lithium niobate crystals that rotated the polarization of an incoming photon by 90~$^{\circ}$ when inactive and preserved the photon's polarization when activated with 200~V. The rise time of this switch was 500~ns. To compensate for any polarization rotations of the delay fibre the optical modulator was set inside a Sagnac loop, similar to the one reported in \cite{sagnac}, to enable  bidirectional operation independent of the input polarization. This strategy was chosen to avoid higher loss associated to polarization maintaining fibre and the need for active polarisation stabilisation.
After switching, the idler photon was incident upon a variable transmission element comprising a half waveplate (HWP) and a PBS, to mimic the transmission of a sample. Since the polarization of the idler photon was mixed after the optical fibre delay, both the horizontal and vertical polarisation components of the idler photon needed to experience the same value of $\eta$---we achieved this by using a beam displacer and half waveplate to convert the two polarisation components of the idler into two path modes with the same polarisation. Both modes then pass through the transmission element and are subsequently focused together onto a free-space APD for detection.

Before any measurements of transmission, we first characterized the performance of the setup. The Klyshko efficiencies of the source without the switch were $\eta_S = 41\%$ and $\eta_I=44\%$ for the signal and idler channels respectively, corrected for dark counts but including $\eta_{\textrm{source}}$ and $\eta_{\textrm{det}}\sim65\%$. After introducing the switch, the efficiency of the idler photon's path (without a sample) was reduced to $\eta_{\textrm{source}}.\eta_{\textrm{det}}= 38\%$, which meant a loss of approximately 15$\%$ in the Sagnac loop and the delay line. The Klyshko efficiency of the signal channel increased to $\eta_S\sim$90$\%$\, which is less than the ideal $\eta_S=100\%$ due to the $\sim1~\mu$s width of the switching window that permits unheralded photons to be leaked through the switch. The pump power was adjusted to minimize this effect having a detection rate in the idler path of $\sim$14 k counts/s. To verify that the source was heralding true single photons we measured the second order correlation function of the idler mode using the triple coincidence method reported in Ref.~\cite{beck2007comparing} obtaining a value of g$^{(2)}(0)=0.031 \pm 0.002$ (where g$^{(2)}(0)=0$ corresponds to perfect single photons and g$^{(2)}(0)=1$ corresponds to Poisson distributed light). We estimated the transmission of the sample $\eta$ as the ratio between $\eta_I$ measured at different sample transmission conditions and $\eta_\text{{source}}.\eta_\textrm{{det}}$, which we characterise by measuring $\eta_I$ with sample transmission set to $\eta = 1$. The statistical precision of the transmission estimate per probe input to the sample is given by the inverse of
\begin{figure}
 \includegraphics[width=1\columnwidth]{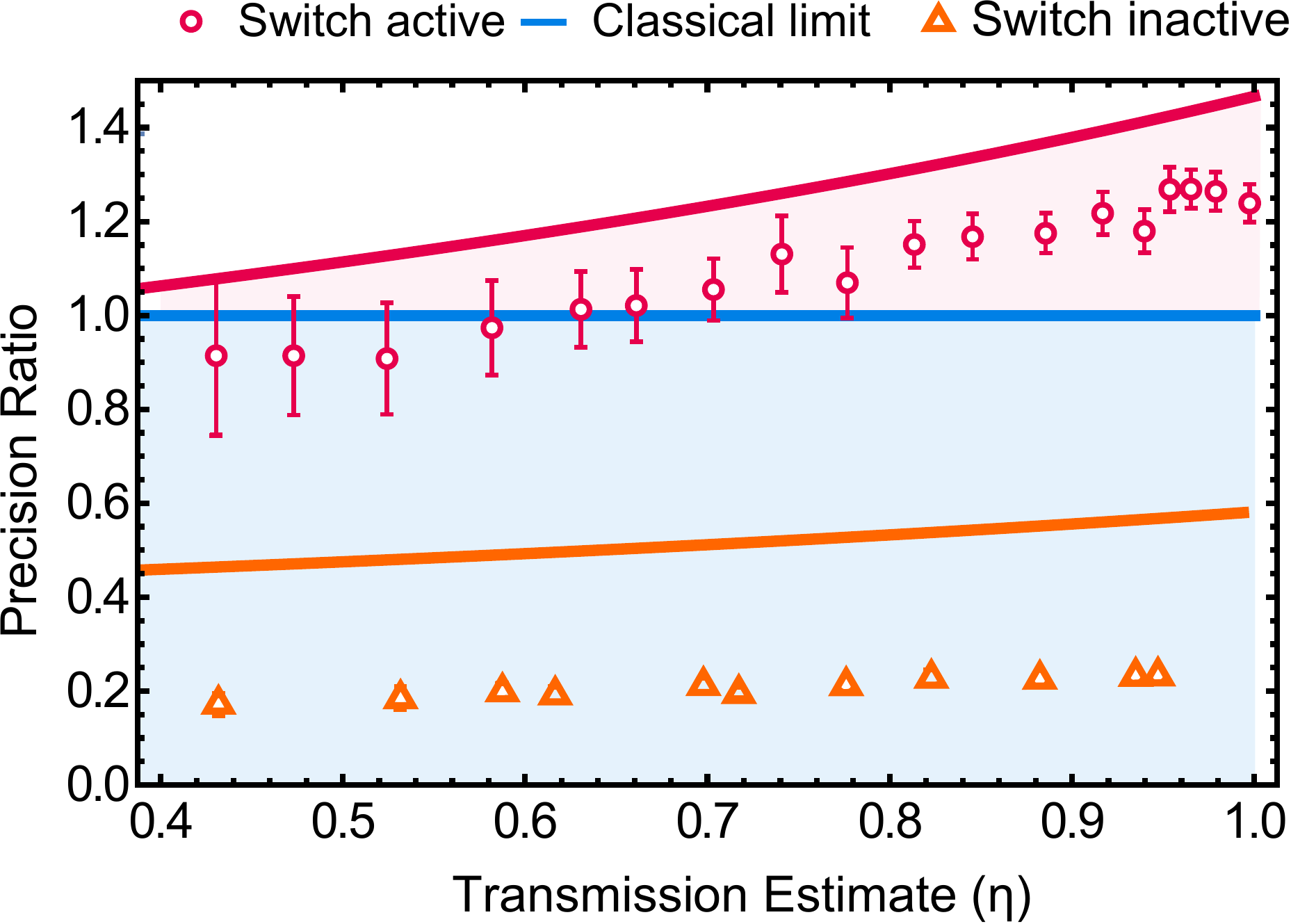}
 \caption{\textbf{Experimental Results.} The pink circles correspond to the estimated experimental advantage compared to a coherent state having the same detector efficiency. Each point corresponds to 3000 repetition of measurements taken with an integration time of 0.2 s and a coincidence window of 30 ns. The orange triangles correspond to the performance of the scheme when the switch is not active (error bars are too small to be seen). The pink solid line corresponds to the expected trend for mixed states with setup efficiency $\eta_\textrm{source}.\eta_\textrm{det}=38\%$ and $1-\eta_S=10\%$ leakage. The orange line corresponds to a mixed state with the same setup efficiency but with $1-\eta_S=62\%$ leaked or unheralded photons. The difference between the solid lines and the experimental are attributed to mechanical noise in fibre coupling efficiency. Error bars were obtained by calculating the variance of binned sets of data points.}
 \label{Adv}
\end{figure}

\begin{equation}
\Delta^2\eta=\text{Var} \left(\frac{\eta_{\text{I}}}{\eta_{\text {source}}}\right)\bar{N}_{\text {probe}},
\label{experimentalFI}
\end{equation}
where $\bar{N}_\text {{probe}}$ is the average number of probe photons given by the number of detected idler photons ($N_I$) corrected for leaked photons through the switch ($\eta_{S}$) absorbed photons by the sample ($\eta$) dark counts of the detector ($N_{D}$) and detector efficiency ($\eta_{\text {det}}$):
\begin{equation}
\bar{N}_\text {{probe}}=\frac{\bar{N}_{I}}{\eta_\text {{det}}.\eta.\eta_{S}}-N_{D}.
\label{N}
\end{equation}

In Fig.~\ref{Adv} we present the precision achievable with our feed-forward transmission measurement setup, with respect to the theoretical precision achievable with a coherent state scheme using the same detector efficiency~\cite{MoreauPrep}. We make this comparison by computing the ratio of precision of the two schemes, as for Fig.~\ref{FIgraph} and we observe a factor of improvement of up to 1.27$\pm.08$ for {$\eta = 0.97$} and a quantum advantage is observable down to sample transmission of {$\eta = 0.65$}. When we turn off the optical switch, the performance of the setup is far below that of the coherent state strategy.

Using feed-forward for measurement is advantageous when it is desired to probe an object with a controlled number of photons~\cite{wo-natphot-7-28,frogexperiment, Vaziri}. Solid state sources of photons, such as quantum dots, could also be used for such purposes. They can operate with MHz emission rates~\cite{hoang2015ultrafast}, they can be used with high heralding efficiency~\cite{ding2016demand} and they can emit higher energy photons~\cite{kimultrafast} than those demonstrated in this letter --- however, the higher specification solid state photon sources currently require additional resources, in particular cryogenic cooling and narrow-band filtering from photonic structure engineering,  that can limit practicality and add cost to development. Practical application of using feed-forward for measurement will be aided by improvements in the brightness of the source~\cite{mccusker2009efficient} and the switching speed.  Increasing the precision obtainable per unit intensity will come with improvements in the loss budget of the setup and increasing detector efficiency. State of the art SPDC sources using superconducting detectors have reported Klyshko efficiencies of 83$\%$ (Ref.~\cite{ramelow2013highly}) --- such an efficiency would already translate into a $\sim5$-fold advantage in precision in our setup. Incorporating the wavelength tunability available in SPDC sources can enable sub shot noise measurement of spectral response~\cite{whittaker2015quantum}. The polarization independent switch used in our experiment could also be useful as the feed-forward mechanism to engineer quantum states that are more complex and have more utility than single photons~\cite{ca-prl-99-163604, mccusker2009efficient}. 
 
We thank M. Loutit and A. Neville for technical assistance. This work was supported by EPSRC, ERC and QUANTIC. JCFM and JGR  acknowledge fellowship support from the Engineering and Physical Sciences Research Council.

\end{document}